\documentclass[12pt]{article}
\usepackage{amsmath,amsfonts,amssymb,amsthm,amstext,amscd,eucal,xcolor}
\usepackage{verbatim}
\usepackage{mathrsfs}
\usepackage[all]{xy}
\usepackage[pagebackref=false]{hyperref}
\usepackage{epsfig}
\usepackage{color}
\usepackage{graphicx}
\usepackage[active]{srcltx}
\usepackage[affil-it]{authblk}
\makeatletter \@addtoreset{equation}{section}

\makeatletter\renewcommand\section{\@startsection {section}{1}{\z@}%
                                   {-3.5ex \@plus -1ex \@minus -.2ex}%nn
                                   {2.3ex \@plus.2ex}%
                                   {\normalfont\large\bfseries}}
\renewcommand\subsection{\@startsection{subsection}{2}{\z@}%
                                     {-3.25ex\@plus -1ex \@minus -.2ex}%
                                     {1.5ex \@plus .2ex}%
                                     {\normalfont\bfseries}}

\begin{document}
\baselineskip 18pt%

\newcommand{\beq}{\begin{equation}}
\newcommand{\eeq}{\end{equation}}
\newcommand{\beqa}{\begin{eqnarray}}
\newcommand{\eeqa}{\end{eqnarray}}
\newcommand{\beqar}{\begin{eqnarray*}}
\newcommand{\eeqar}{\end{eqnarray*}}
\renewcommand{\a}{\alpha}
\renewcommand{\b}{\beta}
\renewcommand{\d}{\delta}
\newcommand{\D}{\Delta}
\newcommand{\e}{\epsilon}
\newcommand{\g}{\gamma}
\newcommand{\G}{\Gamma}
\newcommand{\h}{\eta}
\newcommand{\ka}{\kappa}
\renewcommand{\l}{\lambda}
\renewcommand{\L}{\Lambda}
\newcommand{\na}{\nabla}
\renewcommand{\O}{\Omega}
\newcommand{\p}{\phi}
\newcommand{\s}{\sigma}
\renewcommand{\t}{\theta}
\newcommand{\z}{\zeta}
\newcommand\cR{{\cal R}}
\newcommand{\al}{\alpha}
\newcommand{\be}{\beta}
\newcommand{\del}{\delta}
\newcommand{\eps}{\epsilon}
\newcommand{\ga}{\gamma}
\newcommand{\Ga}{\Gamma}
\newcommand{\inn}{\!\cdot\!}
\newcommand{\kk}{\varphi}
\newcommand\F{{}_3F_2}
\newcommand{\la}{\lambda}
\newcommand{\La}{\Lambda}
\newcommand{\Om}{\Omega}
\newcommand{\sig}{\sigma}
\renewcommand{\t}{\theta}
\newcommand{\ssc}{\scriptscriptstyle}
\newcommand{\eg}{{\it e.g.,}\ }
\newcommand{\ie}{{\it i.e.,}\ }
\newcommand{\labell}[1]{\label{#1}} %{\label{#1}} %
\newcommand{\labels}[1]{\label{#1}}%{\vskip-2ex$_{#1}$\label{#1}}
\newcommand{\reef}[1]{(\ref{#1})}
\newcommand\prt{\partial}
\newcommand\veps{\varepsilon}
\newcommand{\pol}{\varepsilon}
\newcommand\ls{\ell_s}
\newcommand\cL{{\cal L}}
\newcommand\cG{{\cal G}}
\newcommand\cI{{\cal I}}
\newcommand\cl{{\iota}}
\newcommand\cP{{\cal P}}
\newcommand\cV{{\cal V}}
\newcommand\cg{{\it g}}
\newcommand\cB{{\cal B}}
\newcommand\cO{{\cal O}}
\newcommand\tcO{{\tilde {{\cal O}}}}
\newcommand\bz{\bar{z}}
\newcommand\bw{\bar{w}}
\newcommand\hF{\hat{F}}
\newcommand\hA{\hat{A}}
\newcommand\hT{\hat{T}}
\newcommand\htau{\hat{\tau}}
\newcommand\hD{\hat{D}}
\newcommand\hf{\hat{f}}
\newcommand\hR{\hat{R}}
\newcommand\hg{\hat{g}}
\newcommand\hp{\hat{\phi}}
\newcommand\hh{\hat{h}}
\newcommand\ha{\hat{a}}
\newcommand\hQ{\hat{Q}}
\newcommand\hP{\hat{\phi}}
\newcommand\hS{\hat{S}}
\newcommand\hX{\hat{X}}
\newcommand\tL{\tilde{\cal L}}
\newcommand\hL{\hat{\cal L}}
\newcommand\tG{{\widetilde G}}
\newcommand\tg{{\widetilde g}}
\newcommand\tphi{{\widetilde \phi}}
\newcommand\tPhi{{\widetilde \phi}}
\newcommand\ta{{\tilde a}}
\newcommand\tb{{\tilde b}}
\newcommand\tc{{\tilde c}}
\newcommand\td{{\tilde d}}
\newcommand\te{{\tilde e}}
\newcommand\tk{{\tilde k}}
\newcommand\ti{{\tilde i}}
\newcommand\tj{{\tilde j}}
\newcommand\tig{{\tilde g}}
\newcommand\tih{{\tilde h}}
\newcommand\tm{{\tilde m}}
\newcommand\tn{{\tilde n}}
\newcommand\tf{{\tilde f}}
\newcommand\tl{{\tilde l}}
\newcommand\tF{{\widetilde F}}
\newcommand\tE{{\widetilde E}}
\newcommand\tpsi{{\tilde \psi}}
\newcommand\tX{{\widetilde X}}
\newcommand\tD{{\widetilde D}}
\newcommand\tO{{\widetilde O}}
\newcommand\tS{{\tilde S}}
\newcommand\tB{{\widetilde B}}
\newcommand\tA{{\widetilde A}}
\newcommand\tT{{\widetilde T}}
\newcommand\tC{{\widetilde C}}
\newcommand\tV{{\widetilde V}}
\newcommand\thF{{\widetilde {\hat {F}}}}
\newcommand\Tr{{\rm Tr}}
\newcommand\tr{{\rm tr}}
\newcommand\M[2]{M^{#1}{}_{#2}}

\makeatletter
\newcommand\xleftrightarrow[2][]{%
	\ext@arrow 9999{\longleftrightarrowfill@}{#1}{#2}}
\newcommand\longleftrightarrowfill@{%
	\arrowfill@\leftarrow\relbar\rightarrow}
\makeatother

\parskip 0.3cm

%\vspace*{1cm}
%\begin{center}
\begin{titlepage}
{\title{\bf{On T-duality  of $R^2$-corrections to DBI  action
at all orders of gauge field}}}
\vspace{.5cm}
\author{Ahmad Ghodsi \thanks{a-ghodsi@ferdowsi.um.ac.ir}}
\author{Mohammad R. Garousi \thanks{garousi@um.ac.ir}}
\author{Ghadir Jafari \thanks{ghadir.jafari@stu-mail.um.ac.ir}}
\vspace{.5cm}
\affil{ Department of Physics, Ferdowsi University of Mashhad,    
\hspace{5.5cm} P.O.Box 1436, Mashhad, Iran}
\renewcommand\Authands{ and }
\maketitle

\begin{abstract}
Recently, it has been observed that in a T-duality invariant world-volume theory in flat spacetime,  all orders of  gauge field strength   and all orders of the D-brane velocity    appear in two specific matrices. Using these two matrices, we construct the world-volume couplings of two massless NSNS states at order $\alpha'^2$ and  all orders of the velocity and the gauge field strength, by requiring them to be invariant under the linear T-duality. The standard extension $F\rightarrow F+P[B]$, then produces all orders of the pull-back of B-field into the action. We   compare the resulting couplings for zero velocity and   gauge field strength, with the $\alpha'^2$ terms of the disk-level S-matrix element of two massless NSNS vertex operators   in the presence of a constant background B-field. We have found an exact agreement.

\end{abstract}
\end{titlepage}

%\newpage 
\section{Introduction}
D-branes are non-perturbative objects in superstring theory which play the central role in exploring  different  aspects of the theory, from  statistical computation of black hole entropy \cite{Strominger:1996sh} to realization of the AdS/CFT correspondence \cite{Maldacena:1997re} or appearance of noncommutative geometry in string theory \cite{Seiberg:1999vs}. These objects  should be described by the supersymmetric extension of the cubic string field theory \cite{Witten:1985cc} which includes massless and all infinite tower of the  massive excitations of the open strings. In the Wilsonian effective action, however, the effect of all massive states appears in the higher derivatives of the massless fields.  

At long wavelength limit, the higher derivative terms can be ignored and   D-branes are completely described by Dirac-Born-Infeld (DBI) action \cite{Leigh:1989jq,Bachas:1995kx} which includes constant metric, B-field and dilaton, as well as  the first derivative of the massless NS fields. A non-Abelian extension for DBI action has been proposed in \cite{Tseytlin:1997csa}. The leading higher derivative corrections to the Abelian DBI action should include acceleration which appears through the second fundamental form $\Omega$ (see \eg \cite{Corley:2001hg}) and the first derivative of the gauge field strength in the NS part, as well as the  first and the second derivatives of the metric, B-field and dilaton in the NSNS part. The first derivative of the metric appears also through $\Omega$ and the second derivatives of the metric appears through the curvature terms. 

The leading higher derivative corrections of the metric have been found in \cite{Bachas:1999um} by requiring the consistency of the effective action with the $O(\alpha'^2)$ terms of the corresponding disk-level scattering amplitude   \cite{Garousi:1996ad,Hashimoto:1996kf}. For totally-geodesic embeddings   of the world-volume in the ambient spacetime ($\Omega=0$), the corrections  to the DBI action in string frame  for vanishing gauge field and B-field   and for a constant dilaton  is  the following action \cite{Bachas:1999um}\footnote{In our index notation  the Greek letters  $(\mu,\nu,\cdots)$ are  the indices of the space-time coordinates, the Latin letters $(a,b,c,\cdots)$ are the world-volume indices and the letters $(i,j,k,\cdots)$ are corresponding to the normal coordinates of the D-brane.}:
 \begin{eqnarray}
S&=& \frac{\pi^2\alpha'^2T_p}{48}\int d^{p+1}x\,e^{-\phi}\sqrt{-\tG}\Big(R_{abcd}R^{abcd}-2\hat{R}_{ab}\hat{R}^{ab}-R_{abij}R^{abij}+2\hat{R}_{ij}\hat{R}^{ij}\Big), \label{BG} 
 \end{eqnarray}
 where $\hat{R}_{ab}=\tilde{G}^{cd}R_{cadb}$, $\hat{R}_{ij}=\tilde{G}^{cd}R_{cidj}$ and  $\tilde{G}_{ab}$ is the pull-back of the bulk metric, $\tilde{G}_{ab}=\prt_aX^\mu\prt_bX^\nu g_{\mu\nu}$. In above equation,  the Riemann curvatures are the projections of  the bulk Riemann tensors into the world-volume or the transverse space.   For example: 
 \begin{equation}
R_{abcd}=\prt_aX^{\a} \prt_bX^{\b} \prt_cX^{\mu} \prt_dX^{\nu} R_{\a\b\mu\nu}\, .
 \end{equation}  
The above action includes all orders of the velocity through the pull-back operator in the static  gauge, \ie   $X^a=\sigma^a$ and  $X^i=\lambda^i$ . The acceleration terms $(\Omega\neq 0)$ have been also found in \cite{Bachas:1999um} (see also \cite{Jalali:2015xca}). The consistency of these couplings with T-duality may include all orders of dilaton, B-field and $F$.
 
 In \cite{Garousi:2009dj,Garousi:2011fc}, it  has been shown that the consistency of the couplings in \reef{BG} at zero velocity with linear T-duality transformation, requires $\nabla H$ and $\nabla\nabla \phi$ couplings to appear as: 
 \begin{eqnarray}
 S&=&\frac{\pi^2\alpha'^2T_p}{48}\int d^{p+1}x\,e^{-\phi}\sqrt{-\tilde{G}}\Big(R_{abcd}R^{abcd}-2\hat{\cal{R}}_{ab}\hat{\cal{R}}^{ab}-R_{abij}R^{abij}+2\hat{\cal{R}}_{ij}\hat{\cal{R}}^{ij}\nonumber\\
 &&\qquad\qquad\quad-\frac{1}{6}\nabla_aH_{ijk}\nabla^aH^{ijk}-\frac{1}{3}\nabla_iH_{abc}\nabla^iH^{abc}+\frac{1}{2}\nabla_aH_{bci}\nabla^aH^{bci}
 \Big)\label{alpha2}\, ,
 \end{eqnarray}
where $\hat{\cal{R}}_{\mu\nu}=\hat{R}_{\mu\nu}+\nabla_{\mu}\nabla_{\nu}\phi$ and $H$ is the field strength of the B-field. The consistency of the action \reef{BG} at zero velocity and non-zero $\Omega$ with nonlinear T-duality should also include $H^4$ couplings in which we are not interested in this paper. Such couplings have been found in \cite{Robbins:2014ara,Garousi:2014oya} for O-planes. In this paper, we are interested  to include all orders of the constant Abelian gauge field strength $F$  in the above action.  This can be done by releasing the assumption in \cite{Garousi:2009dj} that the velocity is zero, because the world-volume transverse scalars transform to the gauge fields  under the T-duality transformations. 

Recently, it has been observed in \cite{Garousi:2015qgr} that in a T-duality invariant world-volume theory in flat spacetime,  all orders of  gauge field strength   and all orders of the D-brane velocity    appear in the following two  matrices:
 \begin{align}
  &{ {G}}^{\mu \nu}={}\prt_aX^{\mu}\prt_bX^{\nu}\  {G}^{ab}\, , \nonumber\\
  &\Theta^{\mu \nu}={}\prt_aX^{\mu}\prt_bX^{\nu}\  \Theta^{ab}\,\labell{GT} ,
  \end{align}
  where $ {G}^{ab}$ and $\Theta^{ab}$  are the symmetric and antisymmetric parts of $(\frac{1}{\eta+\prt\lambda^i\prt\lambda_i+F})^{ab}$, respectively. They transform into each other under T-duality \cite{Garousi:2015qgr}.
Using these matrices, in the next section we construct world-volume $R^2, R\phi, \phi^2$, $R\nabla H, \phi H$ and $(\nabla H)^2$ couplings   by requiring them to be invariant under the linear T-duality and to reduce to \reef{BG} when $F=0$. In section three, in order to confirm our result with the S-matrix element, we use the standard extension $F\rightarrow F+P[B]$, to include all pull-back of B-field into the action. We then   compare the resulting couplings for zero velocity and   gauge field strength with the $\alpha'^2$ terms of the disk-level S-matrix element of two massless NSNS vertex operators   in the presence of constant background B-field \cite{Garousi:1998bj,Wijnholt:2003pw}. We will show an exact agreement.

\section{T-duality constraint}

In this section, we are going to impose the constraint that the effective action should be invariant under linear T-duality transformations, to include constant gauge field strength into the action \reef{alpha2}.  Let us  first review the strategy   presented in \cite{Garousi:2009dj} and \cite{Garousi:2013gea}. As we know under T-duality the Neumann and Dirichlet boundary conditions are exchanging \cite{Becker:2007zj}, this is due to the fact that a compact direction  that is  lying in the world-volume of the  D-brane in original picture, becomes a transverse direction  after T-duality. Symbolically if we consider  $S$ as an action for  D-brane and 
  $y$ as a compact direction then under T-duality we will have:  
\begin{equation}
S_{D_{p}}[(a,y),i]  \xleftrightarrow{\text{T-duality}} S_{D_{p-1}}[a,(i,y)]\, ,	\label{T-duality}
\end{equation} 
where $a$ represents the tangent coordinate to the D-brane and $i$ stands for normal coordinate. The Killing index $y$ on the left-hand side is a tangent coordinate whereas on the right-hand side it is a normal coordinate.

The T-duality transformation for gauge field is  $\tilde{A}_y=\lambda^y$ which is a linear transformation. However, the T-duality transformation for NSNS fields are nonlinear \cite{TB,Meessen:1998qm,Bergshoeff:1995as,Bergshoeff:1996ui,Hassan:1999bv}. Assuming NSNS fields as perturbations around the flat space, one can find the  relevant  linear T-duality transformations for the closed string fields, \ie 
\beqa
&&\tilde{\phi}=\phi-\frac{1}{2}{h}_{yy}\,, \quad \tilde{h}_{yy}=-h_{yy}\,, \quad \tilde{h}_{\mu\nu}=h_{\mu\nu}\,,\nonumber \\
&&\tilde{h}_{\mu y}=B_{\mu y}\,,  \quad \tilde{B}_{\mu y}=h_{\mu y}\,,  \quad \tilde{B}_{\mu\nu}=B_{\mu\nu} \,. \label{Tdual}
\eeqa
%Moreover, for simplicity we restrict our calculations in the NSNS part to the metric and B-fields. Since dilaton transforms to %$h_{yy}$, the consistent truncation is to assume $\phi=h_{yy}=0$.   
One can use the T-duality   implied by (\ref{T-duality}) as a constraint to find some new couplings from the known ones, in fact this is the method have been used in \cite{Garousi:2009dj} and \cite{Garousi:2013gea}.  

Before going through the T-duality consideration we note that the curvature terms, and those terms with $\nabla\nabla\phi$ and $\nabla H$ in equation (\ref{alpha2}) are the projections of  the corresponding bulk   tensors into the world-volume or the transverse space. By using the projection operator onto the vector space  normal to the D-brane  defined as:
 \begin{equation}
 \bot^{\mu\nu}=n^\mu_in^{\mu i}=G^{\mu\nu}-\tilde{G}^{\mu\nu}\, ,
 \end{equation}
 where $\tilde{G}^{\mu\nu}=\prt_aX^\mu\prt_bX^\nu \tilde{G}^{ab} $  is the first fundamental form, one can transform  the transverse indices in the action (\ref{alpha2}) into the  bulk  space-time coordinates. 
As an example:  
  \begin{eqnarray}
\hat{R}_{ij}\hat{R}^{ij}&=&\bot^{\mu\nu}\bot^{\a\b} \hat{R}_{\a\mu}\hat{R}_{\b\nu}=\bot^{\mu\nu}\bot^{\a\b}\tilde{G}^{ab} \prt_aX^{\rho} \prt_bX^{\lambda} R_{\a\rho\mu\lambda} \tilde{G}^{cd} \prt_cX^{\sigma} \prt_dX^{\gamma} R_{\b\sigma\nu\gamma}\nonumber\\&=&\bot^{\mu\nu}\bot^{\a\b}\tilde{G}^{\rho\lambda}\tilde{G}^{\sigma\gamma}R_{\a\rho\mu\lambda}R_{\b\sigma\nu\gamma} \,.
  \end{eqnarray}
 Apart from the overall constant factor,  the  Lagrangian density in the action (\ref{alpha2}) can be rewritten  as:
 \begin{eqnarray}
&& \!\!\!\!\!\!{\cal L}=\tilde{G}^{\alpha \beta} \tilde{G}^{\nu \mu}\big(2 
 R_{\alpha}{}^{\rho}{}_{\beta}{}^{\kappa} R_ {\nu \rho \mu \kappa}- R_{\alpha \nu}{}^{\rho \kappa} R_ {\beta \mu \rho 
 	\kappa}   
 \big) -2 \tilde{G}^{\alpha \beta} \tilde{G}^{\nu \mu} \tilde{G}^{\rho \kappa} \big(2 R_{\alpha \nu \beta}{}^{\lambda} R_ {\rho\mu \kappa 
 	\lambda}- 
 R_{\alpha \nu \rho}{}^{\lambda} R_ {\beta \mu \kappa \lambda}  \big) \nonumber\\
 &&\!\!\!\!\!\!
-\tfrac{1}{3} \tilde{G}^{\alpha \beta} 
 \nabla_{\alpha}{H}^{\gamma \delta \rho} \
 \nabla_{\beta}{H}_{\gamma \delta \rho} + \tilde{G}^{\alpha 
 	\beta} \tilde{G}^{\gamma \delta} 
 \nabla_{\gamma}{H}_{\alpha}{}^{\rho \kappa} 
 \nabla_{\delta} {H}_{\beta \rho \kappa}   -\tfrac{2}{3} \tilde{G}^{\alpha \beta} \tilde{G}^{\gamma \delta} \tilde{G}^{\rho 
 	\kappa} \nabla_{\lambda}{H}_ {\beta \delta \kappa} 
 \nabla^{\lambda}{H}_{\alpha \gamma \rho} \label{R2H2}\nonumber \\
&&\!\!\!\!\!\!+ 4 R_{\alpha \lambda \beta \mu}\tilde{G}^{\alpha \beta} \nabla^{\mu}\nabla^{\lambda}\phi - 8 R_{ \lambda\beta \mu \nu}\tilde{G}^{\alpha \beta}\tilde{G}^{\lambda \mu} \nabla^{\nu}\nabla_{\alpha}\phi+2\nabla_{\beta}\nabla_{\alpha}\phi \nabla^{\beta}\nabla^{\alpha}\phi - 4\tilde{G}^{\alpha \beta} \nabla_{\mu}\nabla_{\beta}\phi \nabla^{\mu}\nabla_{\alpha}\phi\,.\nonumber\\
&&
\end{eqnarray}
 Writing  in this form, all the Riemann curvatures, $\nabla H$ and $\nabla\nabla\phi$   will be the bulk tensors. Moreover, in this form, the  transverse scalar fields appear in the action just through the first fundamental form $\tilde{G}^{\mu\nu}$. To find the scalar couplings, it is sufficient to break the indices of $\tilde{G}^{\mu\nu}$ to tangent and normal indices and to go to the static gauge \ie $X^a=\sigma^a$ and  $X^i=\lambda^i$ as follows:
 \begin{align}
  &\tilde{G}^{ia}={}\tilde{G}^{ab}\prt_b\lambda^i\, ,  \nonumber\\
 &\tilde{G}^{ij}={}\tilde{G}^{ab}\prt_a \lambda^i\prt_b\lambda^j\, , \label{SplitG}
 \end{align}
   where $\tilde{G}^{ab}$ is the inverse of the pull-back metric in the static gauge \ie $\tilde{G}^{ab}=(\frac{1}{g +\prt\lambda^i\prt\lambda_i})^{ab}$.
  
The  bulk metric in  $\tilde{G}^{\mu\nu}$ is not flat, so the  Lagrangian density	\reef{R2H2} is covariant and contains all orders of the D-brane velocity. It would be desirable to include all orders of gauge field in a  covariant expression. That expression may be found by expanding  $\tilde{G}^{\mu\nu}$ in \reef{R2H2} in terms of different orders of  $g$ and velocity, and then include appropriate B-field and $F$ to make each term to be invariant under the T-duality.  To restrict the calculation for finding only $F$-terms, we assume the bulk metric in   $\tilde{G}^{\mu\nu}$ to be flat metric.  Then the extension $\tilde{G}^{\mu\nu}\rightarrow G^{\mu\nu}$ where $G^{\mu\nu}$ is given in \reef{GT}, produces  the following couplings: 
 \begin{eqnarray}
 && \!\!\!\!\!\!{\cal L}={G}^{\alpha \beta} {G}^{\nu \mu}( 2 
 R_{\alpha}{}^{\rho}{}_{\beta}{}^{\kappa} R_ {\nu \rho \mu \kappa}- R_{\alpha \nu}{}^{\rho \kappa} R_ {\beta \mu \rho 
 	\kappa}   
 ) -2{G}^{\alpha \beta} {G}^{\nu \mu} {G}^{\rho \kappa}\big(2 R_{\alpha \nu \beta}{}^{\lambda} R_ {\rho\mu \kappa 
 	\lambda}- 
 R_{\alpha \nu \rho}{}^{\lambda} R_ {\beta \mu \kappa \lambda}  \big) \nonumber\\
 && \!\!\!\!\!\!
 -\tfrac{1}{3} {G}^{\alpha \beta} 
 \nabla_{\alpha}{H}^{\gamma \delta \rho} \
 \nabla_{\beta}{H}_{\gamma \delta \rho} + {G}^{\alpha 
 	\beta} {G}^{\gamma \delta} 
 \nabla_{\gamma}{H}_{\alpha}{}^{\rho \kappa} 
 \nabla_{\delta} {H}_{\beta \rho \kappa}   -\tfrac{2}{3} {G}^{\alpha \beta} {G}^{\gamma \delta} {G}^{\rho 
 	\kappa} \nabla_{\lambda}{H}_ {\beta \delta \kappa} 
 \nabla^{\lambda}{H}_{\alpha \gamma \rho} \label{R2H2S} \nonumber \\
&& \!\!\!\!\!\! +4 R_{\alpha \lambda \beta \mu}{G}^{\alpha \beta} \nabla^{\mu}\nabla^{\lambda}\phi - 8 R_{ \lambda\beta \mu \nu}{G}^{\alpha \beta}{G}^{\lambda \mu} \nabla^{\nu}\nabla_{\alpha}\phi+2\nabla_{\beta}\nabla_{\alpha}\phi \nabla^{\beta}\nabla^{\alpha}\phi - 4{G}^{\alpha \beta} \nabla_{\mu}\nabla_{\beta}\phi \nabla^{\mu}\nabla_{\alpha}\phi\,.\nonumber \\
&&
 \end{eqnarray}
Obviously this action reduces to \reef{R2H2} when $F\rightarrow 0$. When the indices of $G^{\mu\nu}$ are not the Killing index $y$, the above expression is invariant under linear T-duality. The matrix $G^{\mu\nu}$ has even number of $F$, so the above couplings are also invariant under the parity.

However, the above action is not invariant under the linear T-duality when the indices of $G^{\mu\nu}$ are   the Killing index. It has been shown in \cite{Garousi:2015qgr} that under T-duality $G^{\mu\nu}$ transforms to $\Theta^{\mu\nu}$. So in the T-duality invariant theory there must be new couplings involving $\Theta^{\mu\nu}$ as well. %For these terms we make an ansatz by imposing a constraint on the couplings to be invariant under parity $(F\to-F\  {\rm \&}\  B\to-B)$. 
Using the fact that $G^{\mu\nu}$ has even number of $F$ and $\Theta^{\mu\nu}$ has odd number of $F$,   the invariance under parity requires   the new  terms contracted by even number of $\Theta$ in $R^2, R\phi $ and $H^2$ terms and odd number in $RH$ and $\phi H$ couplings.\footnote{In fact we have observed that for example by considering a term such as $RR\Theta$, the T-duality transformation leads to $RH{G}$ couplings, which reduce to $RH$ couplings in vanishing $F$ limit, but such terms do not exist in  (\ref{R2H2}), so  T-duality requires parity invariance.} Moreover, the couplings $R^2 $ and $H^2$ in    (\ref{R2H2S}) in which the indices of  $G^{\mu\nu}$ can be the Killing index $y$, contain   two or three  $G^{\mu\nu}$s. As a result, T-duality requires the new terms with structure $R^2$ and $ H^2 $   to have two or three $\Theta$ and/or $G$. The couplings $R\phi$ in which the indices of  $G^{\mu\nu}$ can be the Killing index,  contain   one or two  $G^{\mu\nu}$s. As a result, T-duality requires the new terms with structure $R\phi $ to have two   $\Theta$s, and the terms with structure $H\phi$ to have one, two or three $\Theta$ and/or  $G$. Note that  the indices of  $G^{\mu\nu}$ in $\phi^2$ term can not be the Killing index, so there is no new coupling with structure $\phi^2$.
%we consider all possible terms that  one or two of their indices are contracted by space-time metric. In fact there are other possible terms with higher possible contractions, one can show  that these terms have not contribution by  T-duality requirement. Moreover for terms with no space-time metric contraction T-duality gives rise to a  Gauss-bonnet combination in the action, where in (\cite{Bachas:1999um}) it has been shown to vanish by different string theory arguments.
 
  All possible independent $R^2$ and $R\phi$ terms are then the  followings:
 \begin{align}
 &\mathcal{L}_1=\Theta^{\alpha \beta} \Theta^{\kappa \lambda} \big(\a_1 R_{\alpha \kappa}{}^{\mu \nu} R_{\beta \lambda \mu \nu}  + \a_2 R_{\alpha}{}^{\mu}{}_{\kappa}{}^{\nu} R_{\beta \mu \lambda \nu} + \a_3 R_{\alpha \beta}{}^{\mu \nu} R_{\kappa \lambda \mu \nu} +\rho  R_{\beta \kappa \lambda \mu}   \nabla^{\mu}\nabla_{\alpha}\phi\big) \label{RTT} \nonumber \\ 
 &  +\Theta^{\alpha \beta} \Theta^{\kappa \lambda} {G}^{\mu \nu}\big( \a_4 R_{\alpha \kappa \mu}{}^{\rho} R_{\beta \lambda \nu \rho}  + \a_5 R_{\alpha \mu \kappa}{}^{\rho} R_{\beta \nu \lambda \rho}   + \a_6 R_{\alpha \beta \mu}{}^{\rho} R_{\kappa \lambda \nu \rho}   + \a_7^{\text{}} R_{\alpha \kappa \beta}{}^{\rho} R_{\lambda \mu \nu \rho} \big)\,,
 \end{align}
 where $\alpha_1,\cdots, \alpha_7$ and $\rho$ are constants. Similarly for $H^2$ terms we have:
  \begin{align}
 & {\cal L}_{2}= \Theta^{\alpha \beta} \Theta^{\gamma \delta} \big( \beta_1\! \nabla\!_{\beta}H_{\delta}{}^{\mu \nu} \nabla\!_{\gamma}H_{\alpha \mu \nu}\!+\! \beta_2  \nabla\!_{\nu}H_{\beta \delta}{}^{\mu} \nabla^{\nu}\! H_{\alpha \gamma \mu}\!+\! \beta_3  \nabla\!_{\gamma}H_{\alpha}{}^{\mu \nu} \nabla\!_{\delta}H_{\beta \mu \nu}\!+\! \beta_4  \nabla\!_{\nu}H_{\gamma \delta}{}^{\mu} \nabla^{\nu}\! H_{\alpha \beta \mu} \big) \nonumber \\ 
  & \!+\! \Theta^{\alpha \beta} \Theta^{\gamma \delta} {G}^{\mu \nu}\big(\beta_5 \nabla\!_{\beta}H_{\delta \nu}{}^{\lambda} \nabla\!_{\gamma}H_{\alpha \mu \lambda} \!+\! \beta_6^{\text{}} \nabla\!_{\gamma}H_{\alpha \mu}{}^{\lambda} \nabla\!_{\delta}H_{\beta \nu \lambda}   \!+\! \beta_7 \nabla\!_{\beta}H_{\alpha \mu}{}^{\lambda} \nabla\!_{\delta}H_{\gamma \nu \lambda} \!+\! \beta_8 \nabla\!_{\mu}H_{\alpha \gamma}{}^{\lambda} \nabla\!_{\nu}H_{\beta \delta \lambda}\nonumber \\ 
  &  \!+\! \beta_9 \nabla\!_{\mu}H_{\alpha \beta}{}^{\lambda} \nabla\!_{\nu}H_{\gamma \delta \lambda}\!+\! \beta_{10} \nabla\!_{\beta}H_{\alpha \gamma}{}^{\lambda} \nabla\!_{\nu}H_{\delta \mu \lambda} \!+\! \beta_{11} \nabla\!_{\gamma}H_{\alpha \beta}{}^{\lambda} \nabla\!_{\nu}H_{\delta \mu \lambda}\!+\! \beta_{12} \nabla\!_{\lambda}H_{\gamma \delta \nu} \nabla^{\lambda}H_{\alpha \beta \mu}   \nonumber \\
	& \!+\! \beta_{13} \nabla\!_{\lambda}H_{\beta \delta \nu} \nabla^{\lambda}H_{\alpha \gamma \mu}\big)\,,\label{H2TT}
  \end{align}
where $\beta_1,\cdots, \beta_{13}$ are unknown coefficients. The  new possible  $RH$ and $\phi H$ couplings are:  
\begin{align}
& {\cal L}_{3}=\Theta^{\alpha \beta}
{G}^{\gamma \delta}\big( \sigma_1^{\text{}} R_{\beta \delta \mu \nu}  \nabla_{\alpha}H_{\gamma}{}^{\mu \nu}   \!+\! \sigma_2^{\text{}} R_{\alpha \beta \mu \nu} \nabla_{\delta}H_{\gamma}{}^{\mu \nu} + \sigma_3^{\text{}} R_{\gamma \mu \delta \nu} \nabla^{\nu}H_{\alpha \beta}{}^{\mu}   + \sigma_4^{\text{}} R_{\beta \mu \delta \nu} \nabla^{\nu}H_{\alpha \gamma}{}^{\mu} \nonumber \\ 
&  + \sigma_5^{\text{}} R_{\beta \delta \mu \nu}  \nabla_{\gamma}H_{\alpha}{}^{\mu \nu} \big)  +\Theta^{\alpha \beta} {G}^{\gamma \delta} {G}^{\mu \nu} \big(\sigma_6^{\text{}} R_{\beta \delta \nu \rho}  \nabla_{\alpha}H_{\gamma \mu}{}^{\rho} + \sigma_7^{\text{}} R_{\delta \mu \nu \rho} \nabla_{\beta}H_{\alpha \gamma}{}^{\rho}    + \sigma_8^{\text{}} R_{\delta \mu \nu \rho} \nabla_{\gamma}H_{\alpha \beta}{}^{\rho}   \nonumber \\
&    + \sigma_9^{\text{}} R_{\beta \mu \nu \rho} \nabla_{\delta}H_{\alpha \gamma}{}^{\rho}   + \sigma_{10}^{\text{}} R_{\alpha \beta \nu \rho} \nabla_{\delta}H_{\gamma \mu}{}^{\rho}  + \sigma_{11}^{\text{}} R_{\beta \delta \nu \rho}\nabla_{\mu}H_{\alpha \gamma}{}^{\rho}   + \sigma_{12}^{\text{}} R_{\beta \nu \delta \rho} \nabla_{\mu}H_{\alpha \gamma}{}^{\rho}\big)   \label{RHTT} \nonumber \\ 
& +\Theta^{\alpha \beta} \Theta^{\gamma \delta} \Theta^{\mu \nu}  \big( \sigma_{13}^{\text{}} R_{\delta \mu \nu \rho} \nabla_{\beta}H_{\alpha \gamma}{}^{\rho} \!+\! \sigma_{14}^{\text{}} R_{\beta \delta \nu \rho} \nabla_{\mu}H_{\alpha \gamma}{}^{\rho}    \!+\! \sigma_{15}^{\text{}} R_ {\beta \delta \nu \rho}\nabla^{\rho}H_{\alpha \gamma \mu}  \!+\! \sigma_{16}^{\text{}} R_ {\delta \mu \nu \rho}\nabla_{\gamma}H_{\alpha \beta}{}^{\rho})
\nonumber\\
&+\gamma_1^{\text{}} \Theta^{\alpha \beta} \Theta^{\gamma \delta} \Theta^{\theta \kappa} \nabla_{\gamma}\nabla_{\alpha}\phi \nabla_{\delta}H_{\beta \theta \kappa}  + \gamma_2^{\text{}} \Theta^{\alpha \beta} \nabla_{\delta}H_{\alpha \beta \gamma} \nabla^{\delta}\nabla^{\gamma}\phi + \gamma_3^{\text{}} \Theta^{\alpha \beta}{G}^{\gamma \delta} \nabla_{\delta}H_{\beta \gamma}{}^{\theta} \nabla_{\theta}\nabla_{\alpha}\phi \nonumber \\ 
 & + \gamma_4^{\text{}} \Theta^{\alpha \beta}{G}^{\gamma \delta} \nabla_{\delta}H_{\alpha \beta}{}^{\theta} \nabla_{\theta}\nabla_{\gamma}\phi  + \gamma_5^{\text{}} \Theta^{\alpha \beta}{G}^{\gamma \delta} \nabla_{\theta}H_{\alpha \beta \delta} \nabla^{\theta}\nabla_{\gamma}\phi\,,
\end{align}  
here again $\sigma_1,\cdots, \sigma_{16}$ and $\gamma_1,\cdots, \gamma_5$ are unknown coefficients. In writing the above terms we have considered independency  by taking care of the Bianchi identities (for more details see appendix A). In this regard (\ref{RTT})-(\ref{RHTT})  are the most general Lagrangians. In our calculation we are going to assume constant $G$ and $\Theta$ and to work with the second order of perturbations, so the terms with coefficients $\beta_3, \beta_6, \sigma_{15}$ and  $\gamma_1$ are total derivatives. Hence,  we ignore them, \ie $\beta_3=\beta_6=\sigma_{15}=\gamma_1=0$. On the other hand,  for some specific relations between the coefficients of some of the above terms, there might be total derivatives which should be dropped. We will find such terms after imposing the T-duality constraint. 

 In order to fix   the unknown coefficients in above Lagrangians by the linear T-duality (\ref{Tdual}), we need to expand the metric around the flat background as $g_{\mu\nu}=\eta_{\mu\nu}+h_{\mu\nu}$ and keep terms up to the second order of perturbation expansion. Moreover one needs to expand   ${G}^{\mu\nu}$ and $\Theta^{\mu\nu}$ up to the third order of velocity and/or  $F$ in the static gauge.
 %\begin{eqnarray}
%&&{G}=(\frac{1}{\tilde{G}+F})_S=\tilde{G}^{-1}+\tilde{G}^{-1}F\tilde{G}^{-1}F\tilde{G}^{-1}+{\it O}(F^4)\, ,\nonumber\\&&
%\Theta=(\frac{1}{\tilde{G}+F})_A=-\tilde{G}^{-1}F\tilde{G}^{-1}-\tilde{G}^{-1}F\tilde{G}^{-1}F\tilde{G}^{-1}F\tilde{G}^{-1}+{\it %O}(F^5)\, .
% \end{eqnarray}
% For scalar fields  we  use the static gauge in (\ref{SplitG}).
Performing these steps for (\ref{R2H2S})-(\ref{RHTT}), we  compute both sides of relation (\ref{T-duality}). Next we apply the linear  T-duality on left(right) hand side and identify it with  right(left) hand side to fix the unknown coefficients.
After all we find the following relations: 
    \begin{eqnarray}
 &&\a_2=0 \ ,\quad \ \a_3=-\frac{\a_1}{2}\ ,\quad \ \a_5=-4\ ,\quad \ \a_6=2-\frac{\a_4}{2}\ ,\quad \ \a_7=8 \ ,\quad \ \rho=-8\,;   
  \nonumber\\&& \b_1=0 \ ,\quad \ \b_2=0 \ ,\quad \ \b_4=1 \ ,\quad \ \b_7=\frac{\b_{10}}{2}+\b_{11}+\b_5\ , \nonumber\\
	&& \b_8=\frac{\b_{10}}{4} \ ,\quad \b_9=\frac{\b_{11}}{4} \ ,\quad \ \b_{12}=-2-\frac{\b_{11}}{4}\ ,\quad \ \b_{13}=2-\frac{\b_{10}}{4}\, ; \label{TdualCs1}
  \\&&\sigma_2=\frac{\sigma_1}{2}\ ,\quad \ \sigma_3=-4 \ ,\quad \ \sigma_4=0 \ ,\quad \ \sigma_5=4 \ ,\quad \ \sigma_6=16+2\sigma_{10}\ , \nonumber\\&&  \sigma_7=\sigma_{11}\ ,\quad \ \sigma_8=-8 \ ,\quad \ \sigma_9=16-\sigma_{11}\ ,\quad \ \sigma_{12}=-16 \ ,\quad \ \sigma_{14}=8 +\frac{\sigma_{13}}{2}\,; \nonumber \\&&
	\gamma_2=2 \ ,\quad \  \gamma_3=2\gamma_4 \ ,\quad \ \gamma_5=-4-\gamma_4  \,. \nonumber
\end{eqnarray} 
In obtaining  the above results we have used the integration by part and assumed $G^{\mu\nu}$ and  $\Theta^{\mu\nu}$ are constant. %also the totally geodesic condition $\Omega=\prt\prt\lambda=0$ and its T-dual equivalent, $\prt F=0$. 
 Substituting (\ref{TdualCs1}) into the Lagrangian (\ref{RTT})  for $R^2$ and $ R\phi$ terms, we have found: 
 \begin{eqnarray}
 &&\mathcal{L}_1= \Theta^{\alpha \beta} \Theta^{\gamma \delta} {G}^{\rho \kappa} \big(- 4 R_{\alpha \rho \gamma}{}^{\lambda} R_{\beta \kappa \delta \lambda}  + 2 R_{\alpha \beta \rho}{}^{\lambda} R_{\gamma \delta \kappa \lambda}   + 8 R_{\alpha \gamma \beta}{}^{\lambda} R_{\delta \rho \kappa \lambda} \big)\nonumber\\&& -8 \Theta^{\alpha \beta} \Theta^{\kappa \lambda} R_{\beta \kappa \lambda \theta} \nabla^{\theta}\nabla_{\alpha}\phi.
 \label{R2ATT}
 \end{eqnarray}
We have also found that  the  unfixed coefficients, $\a_1$ and $\a_4$ appear in the following expressions: 
 \begin{align}
 \a_1^{\text{}} \Theta^{\alpha \beta} \Theta^{\gamma \delta} (R_{\alpha \gamma}{}^{\rho \kappa} R_{\beta \delta \rho \kappa}  -  \tfrac{1}{2} R_{\alpha \beta}{}^{\rho \kappa} R_{\gamma \delta \rho \kappa} )
  + \a_4^{\text{}}\Theta^{\alpha \beta} \Theta^{\gamma \delta} {G}^{\rho \kappa} (R_{\alpha \gamma \rho}{}^{\lambda} R_{\beta \delta \kappa \lambda}  -  \tfrac{1}{2} R_{\alpha \beta \rho}{}^{\lambda} R_{\gamma \delta \kappa \lambda} )\,.\nonumber
\end{align}
However, they  are   total derivatives up to the second order of the  perturbations. Since our calculations are valid up to the second order of the perturbations, we can ignore these terms, \ie $\a_1=\a_4=0$. %and the T-duality relation (\ref{T-duality}) is correct  up to total derivatives, therefore these coefficients can't be fixed by this method. 
  For $H^2$ terms, we have found
 \begin{eqnarray}
 \mathcal{L}_2=\Theta^{\alpha \beta} \Theta^{\mu \nu} \nabla_{\sigma}H_{\mu \nu}{}^{\rho} 
 \nabla^{\sigma}H_{\alpha \beta \rho}+2\Theta^{\alpha \beta} \Theta^{\mu \nu} 
 {G}^{\rho \sigma}\big(\nabla_{\lambda}H_{\beta \nu \sigma} 
 \nabla^{\lambda}H_{\alpha \mu \rho} - 
 \nabla_{\lambda}H_{\mu \nu \sigma} \nabla^{\lambda}H_{\alpha 
 	\beta \rho}   \big).    \label{H2ATT}
 \end{eqnarray}
We have also showed that the constants $\beta_5, \beta_{10}$ and $\beta_{11}$ are the coefficients of total derivative terms, so we set them to zero. % where $\mathcal{L}_2(Amb)$ are ambiguous terms presented in equation (\ref{AmbH2TT}) in Appendix B which are again total derivatives  and their coefficients can't be fixed. 
 Finally for $RH$ and $\phi H$ couplings in (\ref{RHTT}) we get:  
  \begin{eqnarray}
&&\!\!\!\!\!\!\mathcal{L}_3=8\Theta^{\alpha \beta} {G}^{\gamma \delta} {G}^{\mu \nu}\big(2 R_{\beta \delta \nu \rho}\nabla_{\alpha}H_{\gamma \mu}{}^{\rho}    -  R_{\delta \mu \nu \rho}\nabla_{\gamma}H_{\alpha \beta}{}^{\rho}    + 2 R_{\beta \mu \nu \rho}\nabla_{\delta}H_{\alpha \gamma}{}^{\rho} - 2 R_{\beta \nu \delta \rho}\nabla_{\mu}H_{\alpha \gamma}{}^{\rho} \big) \nonumber \\ 
 &&\!\!\!\!\!\!+4\Theta^{\alpha \beta} {G}^{\gamma \delta}\big(  R_{\beta \delta \mu \nu}\nabla_{\gamma}H_{\alpha}{}^{\mu \nu}  \!-\!  R_{\gamma \mu \delta \nu}\nabla^{\nu}H_{\alpha \beta}{}^{\mu} \big) 
   +8 \Theta^{\alpha \beta} \Theta^{\gamma \delta} \Theta^{\mu \nu} \big( R_{\delta \mu \nu \rho}\nabla_{\gamma}H_{\alpha \beta}{}^{\rho} \!+\! R_{\beta \delta \nu \rho}\nabla_{\mu}H_{\alpha \gamma}{}^{\rho} \big) \nonumber\\
		&&\!\!\!\!\!\! +2 \Theta^{\alpha \beta} \nabla_{\delta}H_{\alpha \beta \gamma} \nabla^{\delta}\nabla^{\gamma}\phi -4\Theta^{\alpha \beta}{G}^{\gamma \delta} \nabla_{\theta}H_{\alpha \beta \delta} \nabla^{\theta}\nabla_{\gamma}\phi\,.
  \label{RAHT}
 \end{eqnarray}
We have also found that the constants $\sigma_1, \sigma_{10},\sigma_{11}, \sigma_{13}$ and  $\gamma_4$ are again the coefficients of total derivative terms, so we set them to zero too.

It is interesting to note that the consistency of the couplings \reef{R2H2} with the linear T-duality could uniquely fix all   orders of constant $F$. They are given in \reef{R2H2S}, \reef{R2ATT}, \reef{H2ATT} and \reef{RAHT}. Similar observation has been made in \cite{Garousi:2015qgr} in making the    world-volume transverse scalar couplings at order $\alpha'$ in the bosonic string theory to be consistent with T-duality. In the next section, we are going to compare the above couplings with the corresponding disk-level S-matrix element.

         \section{Comparison with   S-matrix  }
				
In the previous section, we have found the couplings of two massless NSNS states at order $\alpha'^2$ to all orders of gauge field which   appear  through $G$ and $\Theta$. It is   known that  the pull-back of B-field can be included into the D-brane effective action via the replacement $F\rightarrow F+P[B]$. This combination is invariant under B-field gauge transformation as    $P[B]\rightarrow P[B]-P[d\Lambda]$, $F\rightarrow F+P[d\Lambda]$. After this replacement, we set the velocity and gauge field strength $F$ to zero. This produces then the couplings of two NSNS states at order $\alpha'^2$ to all orders of constant B-field. Such couplings may be compared with the disk-level S-matrix element of two NSNS vertex operators in the presence of background B-field.

  %   Considering a constant background field  leads to the following boundary condition for wordsheet along the D-brane directions:
   %  \begin{equation}
   %  \eta_{ab}\prt_nX^a+iB_{ab}\prt_tX^a\mid_{\prt\Sigma}=0\, ,
   %  \end{equation}
    % $\prt\Sigma$ is the wordsheet boundary and $\prt_t$ is  tangential and $\prt_n$ is normal derivative to this boundary. As we see in presence of B-field we have a mixed Neumann and Dirichlet boundary condition. Moreover  we have B-dependent for correlations functions  and therefore vertex operators. Using these vertex operators one can find  
    The S-matrix element of two NSNS vertex operators in the presence of background B-field has been calculated in \cite{Garousi:1998bj,Wijnholt:2003pw}, \ie  
     \begin{equation}
     A=-\frac{ T_p\sqrt{-\det(\eta+B)}}{2}\,
     \frac{\Gamma(-t/2)\Gamma(-2s)}{\Gamma(1-t/2-2s)}
     \left(-2s\,a_1+\frac{t}{2}\,a_2\right) \, ,
     \label{finone}
     \end{equation}
     where $t=-2p_1\inn p_2$ is the square  of the momentum transfer in the transverse directions 
     and $s=-{\frac12}p_1\inn D\inn p_1$ is the momentum flow  parallel to
     the world-volume of the D-brane. $D$ is defined such that for world-volume coordinates it is $D^{ab}=(\frac{\eta+B}{\eta-B})^{ab}$ and for transverse directions it is $D^{ij}=-\delta^{ij}$.
     The kinematic factors $a_1$ and $a_2$  are given by:
     \begin{eqnarray}
         a_1&=&{ Tr}(\pol_1\inn D)\,p_1\inn \pol_2 \inn p_1 -p_1\inn\pol_2\inn
     D\inn\pol_1\inn p_2 - p_1\inn\pol_2\inn\pol_1^T \inn D^T\inn p_1
     -p_1\inn\pol^T_2\inn\pol_1\inn D\inn p_1
     \nonumber\\
     &&-\frac{1}{2} (p_1\inn\pol_2\inn \pol^T_1\inn p_2+p_2\inn\pol^T_1\inn\pol_2
     \inn p_1)-
     s\,{\rm Tr}(\pol_1\inn\pol_2^T)
     +\Big\{1\longleftrightarrow 2\Big\}\, ,
        \nonumber\\
     a_2&=&{\rm Tr}(\pol_1\inn D)\,(p_1\inn\pol_2\inn D\inn p_2 + p_2\inn
     D\inn\pol_2\inn p_1 +p_2\inn D\inn\pol_2\inn D\inn p_2)
     +p_1\inn D\inn\pol_1\inn D\inn\pol_2\inn D\inn p_2
     \nonumber\\
     &&-\frac{1}{2}(p_2\inn
     D\inn\pol_2\inn\pol_1^T\inn D^T\inn p_1+p_1\inn D^T\inn\pol^T_1\inn\pol_2\inn
     D\inn p_2)
     -s\,{\rm Tr}(\pol_1\inn D\inn \pol_2\inn D)
     \nonumber\\
     &&+s\,{\rm Tr}(\pol_1\inn\pol_2^T)
     +{\rm Tr}(\pol_1\inn D) {\rm Tr}(\pol_2\inn D)\,(s+t/4)
     +\Big\{1\longleftrightarrow 2 \Big\}\ \,,
     \label{finthree}\nonumber
\end{eqnarray}
where $\pol_{1}$ and $\pol_{2}$ are the polarizations of the NSNS states. In order to find the corresponding effective action at order $\alpha'^2$,  we need to expand it in powers  of $\a' p^2$ which is given as  
\beq
A=-\frac{T_p\sqrt{-\det(\eta+B)}}{2}  (-2s\,a_1+\frac{t}{2}\,a_2)\left(\frac{1}{st}+\frac{\pi^2\alpha'^2}{24}+O(\alpha'^4)\right)\, . \label{Amp}
\eeq
The leading order of this amplitude contains a t-channel   and s-channel in addition to some contact terms. This order completely is described by supergravity action in the bulk plus the DBI  action on the  D-brane \cite{Garousi:1998bj}\footnote{This is similar to bosonic string computations \cite{Ardalan:2002qt}.}. At  all other   orders, the   amplitude just contains contact terms which are the effective couplings in the momentum space. We are interested in the $\alpha'^2$-contact terms.  

 In what follows we impose physical conditions for graviton ($\pol_{\mu\nu}=\pol_{\nu\mu}$) and B-field  ($\pol_{\mu\nu}=-\pol_{\nu\mu}$)   to find two-graviton, two-B-field and one-graviton-one-B-field couplings   in the presences of background B-field. For two gravitons  we can simplify $O(\a')^2$ part as follows: 
\begin{eqnarray}
A &\sim &s^2 \Tr(\pol_1 \inn \pol_2) -2 s\ \Tr(\pol_1 \inn V_S) p_1\inn \pol_2\inn  p_1   + \tfrac{t}{4} \Tr(\pol_1 \inn  V_S) Tr(\pol_2 \inn V_S)  (4 s + t)\nonumber\\
&& +s t\Tr(\pol_1 \inn \pol_2 \inn V_S)+2 t\ p_1 \inn V_S \inn \pol_1 \inn \pol_2 \inn V_S \inn p_1+2 s\  p_1 \inn \pol_2 \inn V_S \inn \pol_1 \inn p_2-s t \Tr(\pol_1 \inn V_S \inn \pol_2 \inn V_S)\nonumber\\
&&+4s  p_1 \inn  \pol_2 \inn \pol_1\inn V_S\inn p_1-2t\Tr(\pol_1\inn V_S) p_1 \inn  \pol_2 \inn V_S\inn p_1 +2 t\Tr(\pol_1\inn V_S) p_1 \inn V_S \inn \pol_2 \inn V_S \inn p_1\nonumber\\&&-2t\  p_1 \inn V_S\inn \pol_2 \inn V_S \inn \pol_1 \inn V_S \inn p_1- s t \Tr(\pol_1\inn V_A \inn \pol_2 \inn V_A)-4t\  p_1 \inn V_A\inn \pol_2 \inn V_A \inn \pol_1 \inn V_S \inn p_1\nonumber\\&&
 +2 t\Tr(\pol_1\inn V_S) p_1 \inn V_A \inn \pol_2 \inn V_A \inn p_1-2t\  p_1 \inn V_A\inn \pol_2 \inn V_S \inn \pol_1 \inn V_A \inn p_1+\Big\{1\longleftrightarrow 2 \Big\}\, , \label{GGAmp}
\end{eqnarray}
where $V_S$ and $V_A$ are the symmetric and antisymmetric parts of  $V=\frac{\eta+D^T}{2}=\frac{1}{\eta+B}$.  The amplitude is invariant under parity  ($V_S\to V_S$,$V_A\to-V_A$ ). 
    We    observe that the B-dependence appears as $\frac{1}{\eta+B}$ which is the same as the $F$-dependence in the T-duality invariant couplings that we have found in the previous section.
       More precisely,  the parts  of   amplitude   (\ref{GGAmp}) that    contain the symmetric matrix $V_S$, coincide with the  $R^2$ terms in (\ref{R2H2S}).  
The other terms containing  the antisymmetric matrix  $V_A$, also completely reproduce (\ref{R2ATT}). This  part of the S-matrix calculations   has been done already in  \cite{Wijnholt:2003pw}. Our results confirm the computations of \cite{Wijnholt:2003pw} after considering some identities for $R^2$ structures (for more details see appendix A). 

 %We should note that this S-matrix calculation  does not fix the ambiguous terms in the previous section  because as stated %previously   these are total derivatives  at  this order of perturbation expansion.

 The contact terms  for two antisymmetric B-fields can be found by imposing the corresponding polarization tensors in  (\ref{finthree}), \ie  
 \begin{eqnarray}
 A &\sim&\tfrac{1}{2}s(4s+t) \Tr(\pol_1\inn \pol_2)\!-\!2s \ p_2 \inn \pol_1\inn\pol_2\inn p_1\!+\! st \Tr(\pol_1\inn V_S\inn \pol_2)\!-\!4s \ p_1\inn\pol_2\inn\pol_1\inn V_S\inn p_1\nonumber\\&&
 -st \ \Tr(\pol_1\inn V_S\inn\pol_2\inn V_S)+2s \ p_2\inn\pol_1\inn V_S\inn\pol_2\inn p_1-2t \ p_1\inn V_S\inn\pol_1\inn V_S\inn \pol_2\inn V_S\inn p_1\nonumber\\&&
   - s t\ \Tr(\pol_1\inn V_A \inn \pol_2 \inn V_A)+2 t\  p_1 \inn V_A\inn \pol_2  \inn \pol_1 \inn V_A \inn p_1
+\tfrac{1}{4} t (2s+t)\Tr(\pol_1\inn V_A)\Tr(\pol_1\inn V_A)\nonumber\\
&&-2 t\ \Tr(\pol_1\inn V_A) p_1 \inn \pol_2 \inn V_A  \inn p_1 -4t \  p_1 \inn V_A\inn \pol_2 \inn V_A \inn \pol_1 \inn V_S \inn p_1
\!-\!4 t\ \Tr(\pol_1\inn V_A) p_1 \inn V_A \inn \pol_2 \inn V_S \inn p_1\nonumber\\
&&
-2t\  p_1 \inn V_A\inn \pol_2 \inn V_S \inn \pol_1 \inn V_A \inn p_1+\Big\{1\longleftrightarrow 2 \Big\}\, .
 \end{eqnarray}
 Again we see that the part without  $V_A$ is in agreement with $H^2$ terms in (\ref{R2H2S}) and  the remaining part is     reproduced  exactly by (\ref{H2ATT}). 

 The amplitude for one graviton and one antisymmetric B-field has no counterpart  at zero background B-field limit because it  has odd number of $V_A$. Imposing physical conditions for graviton and B-field in amplitude  (\ref{Amp}), we have found
     \begin{eqnarray}
  A &\!\!\!\!\!\sim\!\!\!\!\!& -\!2 s t\ \Tr(\pol_1 \inn \pol_2 \inn V_A)+4 s t\ \Tr(\pol_1 \inn V_S \inn \pol_2 \inn V_A)
  -\tfrac{1}{2} t (4s+t)\ \Tr(\pol_1\inn V_S)\Tr(\pol_1\inn V_A)\nonumber\\&&
    + 4s \  p_1 \inn V_A\inn \pol_2  \inn \pol_1  \inn p_2
      + 4s \  p_1 \inn V_A\inn \pol_1  \inn \pol_2 
      - 4s \  p_1 \inn \pol_2  \inn V_A\inn \pol_1  \inn p_2
      +2s \  p_2 \inn \pol_1 \inn p_2  \Tr(\pol_1\inn V_A) \nonumber\\&&
      -2 t\  p_1 \inn V_S\inn \pol_1\inn p_2  \Tr(\pol_2\inn V_A)
      -4 t\  p_1 \inn V_A\inn \pol_2  \inn \pol_1 \inn V_S \inn p_1
       +2 t\  p_1 \inn V_A\inn \pol_2\inn p_1  \Tr(\pol_1\inn V_S)
      \nonumber\\&& 
  -2 t\  p_1 \inn V_S\inn \pol_1\inn V_S\inn p_1  \Tr(\pol_2\inn V_A) 
\!+\!4 t\  p_1 \inn V_S\inn \pol_1  \inn V_A\inn \pol_2 \inn V_S \inn p_1
 \!-\!4t \  p_1 \inn V_A\inn \pol_2\inn V_S\inn p_1  \Tr(\pol_1\inn V_S)
\nonumber\\&&+4 t\  p_1 \inn V_S\inn \pol_1  \inn V_S\inn \pol_2 \inn V_A \inn p_1-4 t\  p_1 \inn V_S\inn \pol_2  \inn V_S\inn \pol_1 \inn V_A \inn p_1    
  \nonumber\\&& 
  +4 t\  p_1 \inn V_A\inn \pol_1 \inn V_A \inn \pol_1 \inn V_A \inn p_1
  -2 t\  p_1 \inn V_A\inn \pol_1 \inn V_A\inn p_1 \Tr(\pol_2 \inn V_A)\, . 
  \end{eqnarray}
In this amplitude  $\pol_1$ is the polarization of graviton and $\pol_2$ belongs to B-field. These contact terms reproduce exactly the couplings (\ref{RAHT}) in the momentum space. We have also replaced the dilaton polarization $\pol_{\mu\nu}=\eta_{\mu\nu}+\ell_{\mu}p_{\nu}+\ell_{\nu}p_{\mu}$ where $\ell$ is an auxiliary vector satisfying $\ell\inn p=1$, in the amplitude and found exact agreement with the corresponding couplings in the previous section. This ends our illustration of precise agreement between the couplings that we have found in the previous section by the linear T-duality calculations and the S-matrix element of two NSNS vertex operators in the presence of background B-field.

{\bf Acknowledgments}:   This work is supported by Ferdowsi University of Mashhad under the grant 3/28071(1392/06/17).

\appendix

\section{On Riemann polynomial identities}

The method we have used here to construct the independent Riemann polynomials is according to work of \cite{Green:2005qr} and using the Mathematica package xAct \cite{Nutma:2013zea} 
which is projecting Riemann tensors onto their  Young tableaux: 
\begin{equation}
R_{\a \b \mu \nu} \rightarrow \frac{1}{3}(2 R_{\a \b \mu \nu}-R_{\a\nu\b\mu} + R_{\a\mu\b\nu})\, .
\end{equation}
For example  it is easy to show that the identity  $2R_{\a \b \mu \nu}R^{\a\mu\b\nu} = R_{\a\b\mu\nu} R^{\a\b\mu\nu}$ holds.
 Similarly there are other many identities for $R^2$ terms contracted with other tensors, which reduce the number of independent terms.
 As another example one may consider the  following two terms  independent:
 \begin{equation}
 x_1 R_{\alpha \delta}{}^{\theta \kappa} R_{\beta 
 	\gamma \theta \kappa} T^{\alpha \beta} T^{
 	\gamma \delta} + x_2^{\text{}} R_{\alpha \delta}{}^{\theta 
 	\kappa} R_{\beta \theta \gamma \kappa} T^{\alpha 
 	\beta} T^{\gamma \delta}\, ,
 \end{equation}
 where T is an arbitrary tensor. But by projecting these terms one gets:
 \begin{eqnarray}
 &&\tfrac{2}{9} (2 x_1 + x_2) R_{\alpha 
 	\mu}{}^{\rho \kappa} R_ {\beta \nu \rho \kappa} 
 T^{\alpha \beta} T^{\nu \mu} + \tfrac{4}{9} 
 (2 x_1 + x_2) R_{\alpha \mu}{}^{\rho \kappa} R_ {\beta \rho \nu \kappa} T^{\alpha \beta} T^{\nu 
 	\mu}\nonumber \\ 
 &&  + \tfrac{1}{9} (2 x_1+ x_2)  R_{\alpha}{}^{\rho}{}_{\mu}{}^{\kappa} R_{\beta \rho \nu \kappa} T^{\alpha \beta} T^{\nu \mu}- \tfrac{1}{9} (2 x_1  + x_2) R_{\alpha}{}^{\rho}{}_{\mu}{}^{\kappa} R_{
 	\beta \kappa \nu \rho} T^{\alpha \beta}  T^{\nu \mu}\, ,
 \end{eqnarray}
 which means that the above terms are not independent and by $x_2=-2x_1$ we have an identity.
 As another example consider the following three terms:
 \begin{equation}
 x_1 R_{\alpha \mu}{}^{\rho \kappa} R_{\beta \nu \rho \kappa} T^{\alpha \beta} T^{\mu \nu} +  x_2 R_{\alpha}{}^{\rho}{}_{\mu}{}^{\kappa} R_{\beta \rho \nu \kappa} T^{\alpha \beta} T^{\mu 
 	\nu} + x_3 R_{\alpha}{}^{\rho}{}_{\mu}{}^{\kappa} 
 R_{\beta \kappa \nu \rho} T^{\alpha \beta} 
 T^{\mu \nu}\, ,
 \end{equation} 
 we can show that we have an identity when $x_2=-2x_1$ and $x_3=2x_1$ . Similarly  one may go  further to find more  independent structures by generalizing above procedure.
  
To find independent couplings for $\nabla H$, one should impose the Bianchi identity $dH=0$. This can be done by  writing $H$ in terms of $B$.    
%??????????????
%It has been shown that D-branes with constant background Neveu-Schwarz B-field lead to a noncommutative gauge field theory in their world-volume \cite{Seiberg:1999vs}. Having action to all orders in gauge field strength  $ F $, one can obtain another  action in the presence of background B-field just  by replacing $ F $ by $ B+F$ everywhere. 
%Therefore in this picture if we have an action with all orders in gauge field strength, then the action with background B-field dependence will be obtained. The opposite is also correct i.e. by using the B-dependent  vertex operators in  string theory one can get an effective action with all orders in F.

%The actions are not covariant because the metric in $G$ and $\Theta$ is flat. On the other hand, we did note consider terms with structure $(\nabla B)^2R$ or  $(\nabla B)^4$ because we assume $\nabla F=0$. Such terms have at least three gravitons which could not be found from the two graviton amplitude. If $\nabla F\neq 0$ then T-duality may found all such terms. However, there is indication that a T-duality invariant couplings which include all order of $F$ may not have a compact expression when it is written in a covariant form

\end{document}